\documentclass[a4paper,11pt]{article}
\usepackage{jheppub} 
\usepackage{times}
\usepackage{amsmath}
\usepackage{amsfonts}
\usepackage{graphicx}
\usepackage{hyperref}
\usepackage{dcolumn}
\usepackage{bm}

\def\beq{\begin{equation}}
\def\eeq{\end{equation}}
\def\bea{\begin{eqnarray}}
\def\eea{\end{eqnarray}}
\def\beqn{\begin{eqnarray}} 
\def\eeqn{\end{eqnarray}}
\def\beeq{\begin{eqnarray}}
\def\eeeq{\end{eqnarray}}

\def\ep{\epsilon}

\def\nn{\nonumber}
\def\Eq#1{Eq.~(\ref{#1})}
\def\ln#1{\mathrm{log}\left(#1\right)}

\def\li#1{\mathrm{Li_2}\left(#1\right)}

\def\qon#1{q_{#1,0}^{(+)}}

\def\res#1{{\rm Res}\left(#1\right)}

\def\qb{\mathbf{q}}
\def\pb{\mathbf{p}}

\def\lb{\boldsymbol{\ell}}

\newcommand\muUV{\mu_{\rm UV}}

\def\uv{{\rm UV}}

\def\gs{g_{\rm S}}
\def\as{\alpha_{\rm S}}
\def\aas{\frac{\as}{4\pi}}

\def\r{{\rm R}}

\def\ii{\imath 0}
\def\MS{{\overline {\rm MS}}}

\def\Se{\widetilde{S}_{\ep}}

\usepackage{xcolor}

\definecolor{ao(english)}{rgb}{0.0, 0.5, 0.0}
\definecolor{americanrose}{rgb}{1.0, 0.01, 0.24}
\definecolor{amethyst}{rgb}{0.6, 0.4, 0.8}
\definecolor{darkorange}{rgb}{1.0, 0.55, 0.0}
\definecolor{applegreen}{rgb}{0.55, 0.71, 0.0}

\def\res#1{{\rm Res} \left(#1\right)}
\def\ad#1{{\cal A}_{\rm D}^{(#1)}}

\def\hd#1{h_{\rm D}\left(#1\right)}
\def\aduv#1{{\cal A}_{\uv}^{(#1)}}

\def\gf#1{g_f^{(#1)}}
\def\gH#1{g_H^{(#1)}}
\def\go#1{g_{\gamma^*}^{(#1)}}
\def\MM#1#2{\overline{|{\cal M}_{#1}^{(#2)}|^2}}
\def\ps#1{\widetilde \Delta_{#1}}

\newcommand{\valencia}{Instituto de F\'{\i}sica Corpuscular, Universitat de Val\`{e}ncia 
-- Consejo Superior de Investigaciones Cient\'{\i}ficas, 
Parc Cient\'{\i}fic, E-46980 Paterna, Valencia, Spain.}
\newcommand{\culiacanA}{Facultad de Ciencias F\'{\i}sico-Matem\'aticas,
Universidad Aut\'onoma de Sinaloa, Ciudad Universitaria, CP 80000 Culiac\'an, Mexico.}

\newcommand{\salamanca}{Departamento de F\'isica Fundamental e IUFFyM, Universidad de Salamanca, 
37008 Salamanca, Spain.}
\newcommand{\liverpool}{University of Liverpool, Liverpool L69 3BX, United Kingdom.}

\title{Vacuum amplitudes and time-like causal unitary in the loop-tree duality}

\author[b]{The LTD Collaboration, 
Selomit Ram\'{\i}rez-Uribe, }
\author[a]{Andr\'es E. Renter\'{\i}a-Olivo,}
\author[a]{David F. Renter\'{\i}a-Estrada,}
\author[a]{Jorge J. Mart\'{\i}nez de Lejarza,} \author[a]{Prasanna K. Dhani,} 
\author[a]{Leandro Cieri,} 
\author[b]{Roger J. Hern\'andez-Pinto,}
\author[c]{German F. R. Sborlini,}
\author[d]{William J. Torres Bobadilla,} 
\author[a]{and Germ\'an Rodrigo}

\affiliation[a]{\valencia}
\affiliation[b]{\culiacanA}
\affiliation[c]{\salamanca}
\affiliation[d]{\liverpool}

\emailAdd{selomitru@uas.edu.mx}
\emailAdd{andres.renteria@ific.uv.es}
\emailAdd{david.renteria@ific.uv.es}
\emailAdd{jormard@ific.uv.es}
\emailAdd{dhani@ific.uv.es}
\emailAdd{leandro.cieri@ific.uv.es}
\emailAdd{roger@uas.edu.mx}
\emailAdd{german.sborlini@usal.es} 
\emailAdd{torres@liverpool.ac.uk}
\emailAdd{german.rodrigo@csic.es}

\abstract{We present the first proof-of-concept application to decay processes at higher perturbative orders of LTD causal unitary, a novel methodology that exploits the causal properties of vacuum amplitudes in the loop-tree duality (LTD) and is directly well-defined in the four physical dimensions of the space-time. The generation of loop- and tree-level contributions to the differential decay rates from a kernel multiloop vacuum amplitude is shown in detail, and explicit expressions are presented for selected processes that are suitable for a lightweight understanding of the method. Specifically, we provide a clear physical interpretation of the local cancellation of soft, collinear and threshold singularities, and of the local renormalisation of ultraviolet singularities. The presentation is illustrated with numerical results that showcase the advantages of the method.}

\begin{document}
\maketitle
\flushbottom

\section{Introduction}
\label{sec:intro}

In a recent paper~\cite{Ramirez-Uribe:2024rjg}, we have proposed a novel representation of differential observables at high-energy colliders that exploits the manifestly causal properties of the loop-tree duality (LTD) at higher perturbative orders and its connections with directed acyclic graph (DAG) configurations in graph theory~\cite{Aguilera-Verdugo:2020set,snowmass2020,Aguilera-Verdugo:2020kzc, Ramirez-Uribe:2020hes,JesusAguilera-Verdugo:2020fsn,Sborlini:2021owe,TorresBobadilla:2021ivx,Bobadilla:2021pvr,deJesusAguilera-Verdugo:2021mvg,Ramirez-Uribe:2021ubp,Benincasa:2021qcb,Kromin:2022txz,Clemente:2022nll,Capatti:2022mly,Rios-Sanchez:2024xtv,Ramirez-Uribe:2024wua}. This approach, dubbed LTD causal unitary, generalises the method of four-dimensional unsubtraction (FDU)~\cite{Hernandez-Pinto:2015ysa,Sborlini:2016gbr,Sborlini:2016hat,Prisco:2020kyb,Driencourt-Mangin:2017gop,Driencourt-Mangin:2019sfl,Driencourt-Mangin:2019aix,Driencourt-Mangin:2019yhu} in which, unlike subtraction methods~\cite{Kunszt:1992tn,Frixione:1995ms,Catani:1996jh,Catani:1996vz,Catani:2000vq,Nagy:2003qn,Weinzierl:2003fx,Frixione:2004is,Kilgore:2004ty,Catani:2007vq,Catani:2013tia,GehrmannDeRidder:2005cm,Gehrmann-DeRidder:2012too,Daleo:2006xa,Somogyi:2006da,Somogyi:2006db,Somogyi:2008fc,Somogyi:2009ri,Bolzoni:2010bt,Gleisberg:2007md,Czakon:2010td,Czakon:2014oma,Bevilacqua:2013iha,NigelGlover:2010kwr,Currie:2013vh,Bonciani:2015sha,Gaunt:2015pea,Boughezal:2010mc,Boughezal:2011jf,Boughezal:2015dra,DelDuca:2016ily,Magnea:2018hab,Magnea:2018ebr,Caola:2019nzf,Buonocore:2019puv,Delto:2019asp,Engel:2019nfw,Asteriadis:2019dte,Cieri:2020ikq,TorresBobadilla:2020ekr,Bertolotti:2022aih,Devoto:2023rpv}, infrared~(IR) and ultraviolet~(UV) singularities are locally cancelled directly between loop and tree contributions at the integrand level, while introducing relevant improvements and new features that facilitate an efficient implementation. 

The central and novel ingredient of LTD causal unitary is a multiloop vacuum amplitude in the LTD representation that depends on~$\Lambda$ loop momenta, $\{\ell_j\}_{j=1,\ldots,\Lambda}$. In LTD causal unitary, the contribution from the $k^{\rm{th}}$-order in perturbation theory to the differential decay rate of a particle of mass $m_a$ is given by
\bea
&& d\Gamma^{(k)}_a = \frac{d\Lambda}{2 m_a} \,  
\sum_{(i_1\cdots i_n a) \in \Sigma} \ad{\Lambda, {\rm R}}(i_1\cdots i_n a) \, {\cal O}_{i_1\cdots i_n} \, \ps{i_1\cdots i_n \bar a}~, 
\label{eq:mastercausal}
\eea
and 
\beq
d\Gamma^{{\rm N}^k{\rm LO}}_a = \sum_{j=0}^k d\Gamma^{(j)}_a~,
\eeq
where $d\Gamma^{{\rm N}^k{\rm LO}}_a$ denotes the differential decay rate up to (next-to)$^k$-leading order, and the integration measure
\beq
d\Lambda = \prod_{j=1}^{\Lambda-1} d\Phi_{\lb_j} = \prod_{j=1}^{\Lambda-1} \mu^{4-d} \frac{d^{d-1} \lb_j}{(2\pi)^{d-1}}~, 
\label{eq:integrationmeasure}
\eeq
is written in terms of the spatial components of $\Lambda-1$ primitive loop momenta, as the spatial components of one of them are fixed by the decaying particle. A detailed derivation of the integration measure from the customary phase space is presented in Appendix~\ref{app:phasespace}. The number of loops of the vacuum amplitude is $\Lambda = L+N-1$, where $N$ is the total number of external particles in LO kinematics, and $L$ is the maximum number of loops that contribute at the $k^{\rm th}$ perturbative order. In \Eq{eq:mastercausal}, $\Sigma$ denotes the set of all the phase-space configurations contributing at the $k^{\rm th}$ order, with $n\in \{m, \ldots, m+k\}$, where $m$ is the number of final-state particles in LO kinematics. For a decay process, $m=N-1$.

The final states with $n$ particles are then generated from residues, called phase-space residues, of a vacuum amplitude $\ad{\Lambda}$ on the on-shell energies $\qon{i_s}=\sqrt{\qb_{i_s}^2+m_{i_s}^2-\ii}$ of the internal propagators, where $\ii$ is the original complex prescription of a Feynman propagator, $\qb_{i_s}$ the spacial components of the four-momentum $q_{i_s}$, and $m_{i_s}$ its mass,
\bea
\ad{\Lambda,\r}(i_1 \cdots i_n a ) &=& \ad{\Lambda}(i_1 \cdots i_n a ) - \aduv{\Lambda}(i_1 \cdots i_n a)~,
\label{eq:res}
\eea
where 
\bea
\ad{\Lambda}(i_1 \cdots i_n a ) &=& {\rm Res} \left(\frac{x_{a}}{2} \, \ad{\Lambda}, \lambda_{i_1\cdots i_n a}\right)~,
\eea
with
\beq
\lambda_{i_1 \cdots i_n a} = \sum_{s=1}^n \qon{i_s} + \qon{a}~.
\eeq
The counterterm $\aduv{\Lambda}(i_1 \cdots i_n a)$ in \Eq{eq:res} implements a local UV renormalisation. The residue at $\lambda_{i_1 \cdots i_n a} = 0$ is obtained by analytically continuing the initial-state on-shell energy, $q_{a,0}^{(+)}=\sqrt{\qb_a^2+m_a^2-\ii}$, to negative values, i.e., $\qon{a}= - p_{a,0}^{(+)}$. We have defined $x_a = 2 \qon{a}$.

The function ${\cal O}_{i_1\cdots i_n}$ defines the observable under consideration by mapping the spatial components of the internal momenta of the dual vacuum amplitude onto the spatial components of the momenta of the external particles. The external on-shell energies are set by the corresponding equation of motion. The default choice ${\cal O}_{i_1\cdots i_n}=1$ gives the total decay rate after integration.

The last factor in \Eq{eq:mastercausal} encodes energy conservation 
\beq
\ps{i_1\cdots i_n \bar a} =  2\pi \,  \delta(\lambda_{i_1\cdots i_n \bar a})~, 
\eeq
with
\beq
\lambda_{i_1\cdots i_n \bar a} = \sum_{s=1}^n \qon{i_s} - p_{a,0}^{(+)}~.
\eeq

The LTD causal unitary representation in~\Eq{eq:mastercausal} involves the sum over all possible phase-space residues of the vacuum amplitude that lead to the desired final states. It is precisely the sum over the phase-space residues that ensures that most of the unique properties of the vacuum amplitude are preserved. Specifically, the vacuum amplitude in LTD is a function of the on-shell energies and is obtained by replacing the Feynman propagators by causal propagators of the form
\beq
\frac{1}{\lambda_{i_1 \cdots i_m}} = \left(\sum_{s=1}^m \qon{i_s} \right)^{-1}~.
\label{eq:causalvacuum}
\eeq
The numerator of the vacuum amplitude is also a function of the on-shell energies and additionally of the internal masses.

Since the real part of the on-shell energies is positive by definition, the causal propagators in \Eq{eq:causalvacuum} cannot become singular. If all the on-shell energies vanish simultaneously for massless particles, the potential soft singularity is screened by the integration measure, which also vanishes in this limit. As a consequence, the vacuum amplitude cannot exhibit soft or collinear singularities, and remarkably threshold singularities are also absent. The absence of singularities in the vacuum amplitude, apart from UV singularities that are accounted for by the local UV counterterm, ensures that the sum over the phase-space residues is also free of soft, collinear and threshold singularities. General proofs of the matching of singularities between phase-space residues have been presented in Ref.~\cite{Ramirez-Uribe:2024rjg}. Therefore, the master expression for the decay rate in \Eq{eq:mastercausal} is well defined in the four physical dimensions of the space-time, since all potential nonintegrable singularities locally cancel out between different phase-space residues.

It is important to stress that LTD causal unitary is conceptually different from using forward-scattering amplitudes to generate final states via unitarity cuts~\cite{Sterman:1978bi,Sterman:1978bj,Ellis:1980wv,Soper:1998ye,Soper:1999xk,Capatti:2020xjc}\footnote{In Refs.~\cite{Sterman:1978bi,Sterman:1978bj}, vacuum polarization diagrams refer to those with two external photons.}. The forward-scattering method, where initial-state particles are on-shell from the outset, and unitarity cuts involve only final-state particles, is closely related to the well-known optical theorem. In contrast, LTD causal unitary deals with off-shell particles in vacuum amplitudes, i.e. amplitudes without external particles, and unitarity cuts involve both initial and final-state particles. All external particles are set on-shell through a LTD's unique property, which ensures this condition when the inverse of the corresponding causal propagator vanishes. In the context of forward-scattering amplitudes, the underlying reason advocated for the cancellation of singularities is unitarity~\cite{Soper:1998ye}.
In LTD causal unitary the cancellation of singularities occurs because we depart from an amplitude, the vacuum amplitude, which is intrinsically IR safe and threshold free.

The purpose of this paper is to provide a first proof-of-concept implementation of LTD causal unitary for physical processes, and to showcase its coherence and  advantages in a detailed manner, especially with respect to the physical interpretation of the contributing components. 

The outline of the paper is as follows. In Section~\ref{sec:higgs}, we present explicit expressions for the phase-space residues that contribute to the decay processes $H \to q \bar q (g)$ and $\gamma^* \to q \bar q (g)$ at LO and NLO. In Section~\ref{sec:localrenormalization}, we construct the corresponding UV counterterms, which implement a local UV renormalisation, and discuss in detail the wave-function renormalisation and the mass renormalisation scheme. In Section~\ref{sec:toy}, we consider a simplified model that has all the main properties of a NNLO calculation and is therefore suitable for presenting LTD causal unitary at NNLO in a lightweight form. In Section~\ref{sec:numeric}, we present a numerical implementation of LTD causal unitary at NLO and NNLO based on the phase-space residues presented in Sections~\ref{sec:higgs} to \ref{sec:toy}, and illustrate the local cancellation of threshold, double- and triple-collinear/quasicollinear singularities. Finally, in Section~\ref{sec:conclusions} we draw our conclusions and future prospects.

\section{Vacuum amplitudes and decay rates at NLO}
\label{sec:higgs}

\begin{figure}[t]
\begin{center}
\includegraphics[scale=0.45]{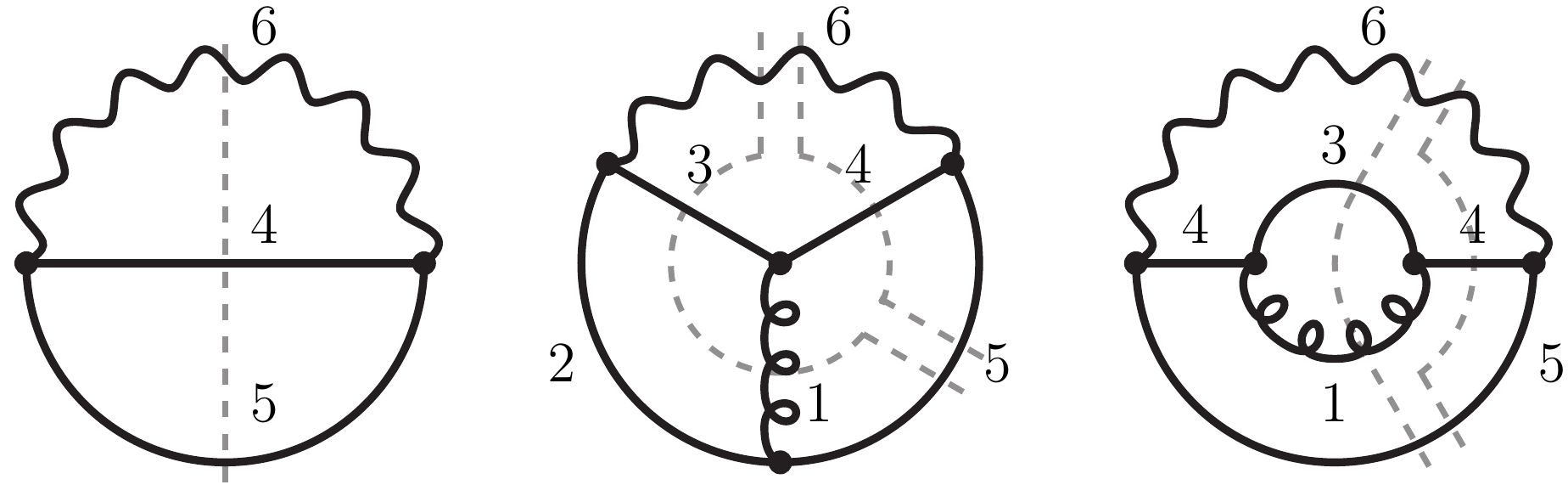}
\caption{The two- and three-loop vacuum diagrams generating $\gamma^*\to q \bar q (g)$ at LO and NLO, respectively. The gray dashed lines represent phase-space residues. Similar diagrams contribute to $H\to q\bar q (g)$ by substituting the photon labeled as $6$ with a Higgs boson.
\label{fig:qqbar}}
\end{center}
\end{figure}

We present explicit expressions up to NLO for the decay processes $H \to q \bar q (g)$ and $\gamma^* \to q \bar q (g)$. These processes are well suited  to easily illustrate all the main features of LTD causal unitary with compact analytical expressions. The kernel vacuum amplitude used to derive the total or differential decay rate, or any other physical observables, is constructed from the vacuum diagrams depicted in Fig.~\ref{fig:qqbar}. 
The two-loop vacuum diagram, on the left, generates the differential expression of the decay rate at LO. The remaining three-loop diagrams contribute to the decay rate at NLO. Note that the diagram on the right and similar diagrams naturally generate selfenergy insertions in external legs. Their contribution is essential to achieve a local cancellation of all IR singularities with the real emission final-state configurations. The internal momenta are labelled as follows in terms of the loop momenta~$\{\ell_s\}_{s=1,2,3}$:
\bea
q_1 = \ell_{12}~, \quad q_2 = \ell_{13}~, \quad q_3 = \ell_1~, \quad
q_4=\ell_2~, \quad q_5 = \ell_{2\bar 3}~, \quad q_6 = \ell_3~.
\label{eq:qs}
\eea
We use the shorthand notation~$\ell_{ij} = \ell_i+\ell_j$ and~$\ell_{i\bar j} = \ell_i-\ell_j$. Nonetheless, it is worth noting that one of the advantages of the LTD representation is that it is independent of the specific labelling of the internal momenta. More precisely, the relationship  between the momenta of the internal propagators and the independent loop momenta in \Eq{eq:qs} can be modified by a linear shift. Yet, the LTD representation in terms of causal propagators and on-shell energies remains unaffexcted by the choice of the specific loop routing. The momentum $q_1$ represents a gluon and is massless, while $q_2$ through $q_5$ are quarks or antiquarks of mass $m$. The momentum $q_6$ is either a Higgs boson or an off-shell photon. To simplify the presentation, we define $x_{i_1\cdots i_n} = \prod_{s=1}^n 2\qon{i_s}$.

The two-loop dual vacuum amplitude associated to the LO decay rate has the compact form 
\beq
\ad{2,f} = \frac{2\, \gf{0}}{x_{456}} \left(  \frac{\MM{f\to q \bar q}{0}}{\lambda_{456}}+ 2 \lambda_{45\bar 6}\right),\,\,f=H,\gamma^*. 
\label{eq:fullAD}
\eeq
The bar above an index indicates that the corresponding on-shell energy bears a minus sign, e.g. $\lambda_{45\bar 6} = \qon{4}+\qon{5}-\qon{6}$. For convenience, the average factor over initial-state polarizations is included in $\ad{2,f}$. In this expression, we have anticipated that $\MM{f\to q \bar q}{0}$ is the matrix element squared obtained with the standard approach of squaring the tree-level amplitude for $f\to q\bar q$ and averaging over the initial-state polarizations, where the interaction couplings and final-state colour factors are collected in the coefficient $\gf{0}$, 
\beq
g_H^{(0)} = y_q^2 C_A~, \qquad g_{\gamma^*}^{(0)} = (e e_q)^2 C_A~,
\label{eq:couplings}
\eeq
with $y_q$ the Yukawa coupling of the quark, $e_q$ the quark electric charge in terms of the electric charge unit $e$, and the Casimir color factor $C_A=3$.

The corresponding phase-space residue is given by 
\beq
\ad{2,f}(456) \equiv
\res{\frac{x_6}{2} \ad{2,f}, \lambda_{456}} = \frac{\gf{0}}{x_{45}} \MM{f\to q \bar q}{0}~,
\eeq
with 
\beq
\MM{H\to q \bar q}{0} = 2 s \beta^2~, \qquad 
\MM{\gamma^* \to q \bar q}{0} = 2 s \left( 1 + \frac{1-\beta^2}{d-2} \right)~, 
\eeq
where $\beta = \sqrt{1-4m^2/s}$ is the velocity of the quark in terms of its mass $m$ and the centre-of-mass energy squared $s$, and $d=4-2\epsilon$ is the number of space-time dimensions in dimensional regularisation (DREG). The factor $1/x_{45}$ is, in effect, the expected factor arising from a phase space as the product of the final-state energies, whereas here it is actually generated in a natural way from the phase-space residue on the vacuum amplitude. The on-shell energy of the decaying particle is $\qon{6} = \sqrt{s-\ii}$, with $s=m_H^2$  for the Higgs boson decay, as we set $\lb_3={\bf 0}$ in the centre-of-mass frame. The partial decay rate at LO is then given by 
\beq
d\Gamma^{\rm LO}_{f\to q\bar q} = \frac{d\Phi_{\lb_2}}{2\sqrt{s}} \, \ad{2,f}(456)  \, \ps{45\bar 6}~. 
\label{eq:decayrateLO}
\eeq
The phase-space condition encoded by $\ps{45\bar 6}$ imposes $\qon{4}=\qon{5}=\qon{6}/2$ in the frame where
$\lb_3={\bf 0}$:
\beq
\ps{45\bar 6} = \frac{\pi}{\beta} \, \delta\left(|\lb_2| - \frac{\beta \sqrt{s}}{2} \right)~.
\eeq
Conservation of the three-momentum is already consistently fulfilled in the vacuum amplitude. Thus, the total decay rates at LO agree with the well-known results
\bea
&& \Gamma^{\rm LO}_{H\to q\bar q}
= g_{H}^{(0)} \frac{\beta^3 m_H}{8\pi}~, 
\qquad \Gamma^{\rm LO}_{\gamma^*\to q\bar q}
= g_{\gamma^*}^{(0)} \frac{\beta \sqrt{s}}{8\pi}
\left( 1 + \frac{2m^2}{s}\right)~. 
\eea
Such a LO calculation, performed through the phase-space residues of a two-loop vacuum amplitude, may seem redundant and overly complex. Yet, the benefits of the method emerge clearly at higher orders.  

We now consider the three-loop vacuum amplitude acting as the kernel dual amplitude at NLO. We shall evaluate phase-space residues involving 3 and 4 external particles, which correspond to the virtual and real contributions to the decay rate at NLO. The differential decay rate at NLO is 
\bea
d\Gamma^{(1)}_{f\to q\bar q} &=& \frac{d\Phi_{\lb_1\lb_2}}{2\sqrt{s}} \, \bigg[
\Big( \ad{3,f, \r}(456) \, \ps{45\bar 6}  + \ad{3,f}(1356) \, \ps{135\bar 6} \Big)  
+ (5\leftrightarrow 2, 4\leftrightarrow 3) \bigg]~.
\label{eq:decayrateNLO}
\eea

The first phase-space residue, unrenormalised, which corresponds to the interference of a one-loop with a tree-level amplitude, is given for the decay of the Higgs boson by 
\bea
&& \ad{3,H}(456) =  
\frac{\gH{1}}{x_{12345}} \bigg[ \bigg\{ \hd{456} \,  \MM{H\to q\bar q}{0} \nn \\
&& \quad + 2(d-2) \lambda_2  \left( - \frac{\lambda_{125} \lambda_{1\bar 2 5}}{\lambda_{13\bar4}} -  
\frac{\lambda_{12\bar 5} \lambda_{1\bar 2 \bar 5}}{\lambda_{134}} + 2\lambda_{1\bar 3}\right) \nn\\ 
&& \quad + \lambda_1 \left( (d-2) \MM{H\to q\bar q}{0} -4 s \right) \left( \frac{1}{\lambda_{23\bar 4 \bar 5}} +\frac{1}{\lambda_{2345}} \right) \bigg\}   + (2\leftrightarrow 3, 4\leftrightarrow 5) \bigg]~, 
\label{eq:a3h}
\eea
and for the decay of an off-shell photon by
\bea
&& \ad{3,\gamma^*}(456)  =  
\frac{\go{1}}{x_{12345}} \bigg[ \bigg\{ \hd{456} \, \MM{\gamma^*\to q\bar q}{0} \nn \\
&& \quad + 2(d-2) \lambda_2  \left( - \frac{\lambda_{125} \lambda_{1\bar 2 5}}{\lambda_{13\bar4}} - 
\frac{\lambda_{12\bar 5} \lambda_{1\bar 2 \bar 5}}{\lambda_{134}} + 2\lambda_{1\bar 3} \frac{d-4}{d-2}\right) \nn\\
&& \quad + \lambda_1 \left( \left( (d-4) \MM{\gamma^*\to q\bar q}{0} -8 s \right) \left( \frac{1}{\lambda_{23\bar 4 \bar 5}} +\frac{1}{\lambda_{2345}} \right)  
\right. \nn \\ 
&& \quad + \left. \frac{4\lambda_{12\bar 5} \lambda_{1 \bar 2 5}}{\lambda_{23\bar 4 \bar 5}} 
+ \frac{4\lambda_{125} \lambda_{1 \bar 2 \bar 5}} {\lambda_{2345}} \right) \bigg\} + (2\leftrightarrow 3, 4\leftrightarrow 5) \bigg], 
\label{eq:a3g}
\eea
where the function $\hd{456}$ is common to both residues,
\bea
&& \hd{456} = s (1+\beta^2)  \left[ \frac{1}{\lambda_{13\bar 4}}  
\left( \frac{1}{\lambda_{23\bar 4 \bar 5}}+ \frac{1}{\lambda_{125}}\right) + \frac{1}{\lambda_{134} \lambda_{2345}} \right]  \nn\\
&& \quad + 4 \lambda_2 \bigg[ \left(\frac{1+\beta^2}{2} + \frac{m^2}{\lambda_{13\bar 4} \lambda_{134}} \right) 
\left(\frac{1}{\lambda_{13\bar 4}} + \frac{1}{\lambda_{134}}\right) 
+ (d-2)\frac{\lambda_{1\bar 3}}{s} \bigg]~. 
\eea 
The interaction coupling at second order is $g_f^{(1)} = \gs^2 \, C_F\, g_f^{(0)}$, where $\gs$ is the strong coupling, $C_F=4/3$ and the tree-level coupling $g_f^{(0)}$ is defined in~\Eq{eq:couplings}. 

The phase-space residues with three particles in the final state are generated from the expression
\bea
&& \ad{3,f}(1356)  = \frac{g_f^{(1)}}{x_{135}} \bigg[ \frac{2}{\lambda_{125} \lambda_{1\bar 2 5} \lambda_{134} \lambda_{13 \bar 4}}  
\bigg( s - \lambda_{125} \lambda_{1\bar 2 5}  \nn \\ && \quad - \lambda_{134} \lambda_{13\bar 4}
- \frac{m^2  (\lambda_{125} \lambda_{1\bar 2 5} +\lambda_{134} \lambda_{13\bar 4})^2 }{\lambda_{125} \lambda_{1\bar 2 5} \lambda_{134} \lambda_{13\bar 4}} \bigg) 
\MM{f\to q\bar q}{0} \nn \\
&& \quad + (d-2) \left( \frac{\lambda_{134} \lambda_{13\bar 4}}{\lambda_{125} \lambda_{1\bar 2 5}}
+ \frac{\lambda_{125} \lambda_{1\bar 2 5} }{\lambda_{134} \lambda_{13\bar 4}} \right) + c_{\rm D}^{(f)} \bigg]~, 
\label{eq:tree1356}
\eea
with $c_{\rm D}^{(H)} = 2(d-2)$, and $c_{\rm D}^{(\gamma^*)} = 2(d-4)$. In this expression the final-state quark is labelled as $3$ and the antiquark as $5$. The exchange of indices in \Eq{eq:decayrateNLO} accounts for the symmetric phase-space residues   
\bea
&&\ad{3,f}(236)  = \left. \ad{3,f}(456)\right|_{(5\leftrightarrow 2, 4 \leftrightarrow 3)}~, \nn \\ 
&&\ad{3,f}(1246)  = \left. \ad{3,f}(1356)\right|_{(5\leftrightarrow 2, 4 \leftrightarrow 3)}~.
\eea

Before moving on to the UV renormalization of the phase-space residues in \Eq{eq:a3h} and \Eq{eq:a3g} let us comment on their IR and threshold singularities. Both expressions become singular at $\lambda_{13\bar4} \to 0$ and this singularity corresponds to a collinear singularity due to the collinear splitting $13\to 4$, where the outgoing quark is labelled as $4$, and the labels $1$ and $3$ denote the internal particles running in the loop. Since the energy of $4$ is limited by energy conservation, the region in which the loop three-momentum $\lb_1$ generates a collinear singularity is bounded~\cite{Buchta:2014dfa}. Another collinear singularity occurs at $\lambda_{12\bar5} \to 0$ when the virtual gluon becomes collinear with the antiquark. If the quark and antiquark are massive, these collinear singularities are shadowed by their mass. Nevertheless, these terms integrate to a mass-dependent logarithm, which is potentially large. The soft singularity, corresponding to a soft virtual gluon, occurs at $\lambda_1\to 0$.

Both soft and collinear singularities, or quasicollinear large logarithms for massive quarks, cancel locally with the tree-level phase-space residues presented in~\Eq{eq:tree1356}, which also diverge as, e.g.,  $1/\lambda_{13\bar4}$ where $1$ and $3$ are now external particles, and $4$ is the parent parton of the collinear splitting. The local cancellation of soft, collinear and quasicollinear configurations ensures that the massless limit of a massive implementation is smooth~\cite{Sborlini:2016hat} and, unlike the state-of-the-art approach, does not require a new calculation.

There is also the expected threshold singularity occurring at $\lambda_{23\bar4\bar5} \to 0$ that generates the absorptive (imaginary) part of the loop amplitude. Threshold singularities are integrable but numerically challenging, and often require the implementation of a contour deformation over the integration domain~\cite{Becker:2010ng,Buchta:2015wna,Capatti:2019edf} or other alternative methods~\cite{Pittau:2021jbs,Kermanschah:2021wbk} to numerically stabilize the integrand. They are known to be one of the main limitations of numerical approaches and imply a compromise between accuracy and lengthy computer jobs. Threshold singularities are, however, absent in the sum of all the unitary phase-space residues because the kernel vacuum amplitude is free of this kind of singularities. This is one of the most remarkable properties of LTD causal unitary.

\section{Local UV renormalisation and mass renormalisation scheme}
\label{sec:localrenormalization}

The phase-space residues with loop remainders are to be renormalised locally by suitable UV counterterms. The novelty introduced in Refs.~\cite{Ramirez-Uribe:2024rjg,Rios-Sanchez:2024xtv} with respect to previous implementations of a local UV renormalisation at one~\cite{Sborlini:2016hat} and two loops~\cite{Driencourt-Mangin:2019aix} (see also~\cite{Becker:2010ng,Donati:2013voa,Cherchiglia:2020iug,Capatti:2022tit}) is to extract the leading UV behaviour directly in LTD by rescaling the on-shell energies:
\beq
\qon{i_s} \to \sqrt{(\rho \, \lb_{s} + \lb_r)^2+m_{i_s}^2 + (\rho^2-1) \mu_{\uv}^2-\ii}~,
\label{eq:uvexpansion}
\eeq
and then by expanding the phase-space residues $\ad{3,f}(456)$ in \Eq{eq:a3h} and \Eq{eq:a3g} for $\rho\to \infty$. In \Eq{eq:uvexpansion}, we assume $\qb_{i_s} = \lb_s + \lb_r$, with $\lb_s$ in the UV region, while $|\lb_r| \ll |\lb_s|$. 

This expansion provides the most singular UV behaviour from which the UV counterterm is constructed. The scale $\mu_\uv$ is interpreted as the renormalisation scale~\cite{Hernandez-Pinto:2015ysa} in the sense that the UV counterterm suppresses all the energy modes of $\ad{3,f}(456)$ above $\mu_\uv$, and locally matches its UV behaviour at very high energies. The UV counterterm renders the phase-space residue UV finite. Yet, it is also necessary to subtract subleading terms to deliver the expected results in, e.g., the $\MS$ renormalisation scheme, after setting the space-time dimensions to $d=4$ in the integrand. It is important to stress that we do not perform the UV expansion diagram by diagram, but consider the vacuum amplitude including all contributions. This approach is more efficient because it takes into account possible cancellations in the UV regions of the loop momenta and leads to more compact local UV counterterms.

The local UV counterterm for the Higgs boson decay reads
\bea
&& \aduv{3,H}(456)  =  \frac{\gH{1}}{x_{45}} \bigg[ \Delta Z_H^{(\uv)} \, \MM{H\to q\bar q}{0} 
- \Delta Z_m^{(\uv)} \, 8 m^2 \left(1+\beta^2 \right) 
+ {\bf \Delta}_{H}^{(\uv)} \bigg]~, 
\label{eq:uvcounterH}
\eea
where 
\bea \label{eq:uvcounterZH}
&& \Delta Z_H^{(\uv)} = \frac{1}{4 \lambda_{\uv}^3} \left( c_{H}^{(\uv)}- c_{\gamma}^{(\uv)} + \frac{3\mu_\uv^2}{2 \lambda_\uv^2}\right)~,  \\
&& \Delta Z_m^{(\uv)} = \left. \frac{1}{4 \lambda_{\uv}^3} \left( c_{H}^{(\uv)}- c_{\gamma}^{(\uv)} + \frac{15\mu_\uv^2}{2 \lambda_\uv^2}\right)\right|_{\mu_\uv=m}~. \nn 
\eea
The coefficients $c_H^{(\uv)}=d$ and $c_\gamma^{(\uv)}=(d-2)/2$ were defined in Ref.~\cite{Driencourt-Mangin:2019aix} and $\lambda_{\uv} = \sqrt{\lb_1^2+\mu_{\uv}^2-\ii}$. The terms proportional to $\mu_\uv^2$ in \Eq{eq:uvcounterZH} are UV subleading contribution that determine the renormalisation scheme. The function 
\bea
&& {\bf \Delta}_H^{(\uv)} = 
\frac{c_{\gamma}^{(\uv)}}{s \lambda_{\uv}^3}  
\left( - \frac{3(\lb_1\cdot \lb_2)^2}{\lambda_{\uv}^2} + \lb_2^2 + 2 \lb_1\cdot \lb_2 \right) 
\MM{H\to q\bar q}{0}~,
\eea
integrates to zero in $d$ space-time dimensions and therefore has no effect, but its contribution is essential to suppress the singular UV angular behaviour at very high energies. 

The integrated UV counterterm is 
\bea
&&\int d\Phi_{\lb_1} \aduv{3,H}(456) \ps{45\bar 6} = \frac{\pi}{s \beta} \, \gH{1} \, \Se
 \bigg[ \frac{\mu^{2\epsilon}}{\mu_\uv^{2\epsilon}}  \frac{3}{\epsilon} \MM{H\to q\bar q}{0} \nn \\ && \quad - \frac{\mu^{2\epsilon}} {m^{2\epsilon}} \left( \frac{3}{\epsilon}  + 4 \right) 8 m^2 (1+\beta^2)  \bigg]~,
\label{eq:integratedUVH}
\eea
where $\Se = (4\pi)^{\epsilon-2} \Gamma(1+\epsilon)$. An important comment is in order here. The on-shell renormalisation scheme is defined by imposing that the renormalised selfenergy and its derivative with respect to the external momentum vanish when the external particle is on shell. The second condition fixes the renormalisation of the wave function, and the first fixes the renormalisation of the mass. The result in \Eq{eq:integratedUVH} is consistent with the customary approach where the term proportional to the squared LO amplitude receives contributions from the UV renormalisation of the $H\to q\bar q$ interaction vertex, and the wave functions of the quark and antiquark. In other words, \Eq{eq:uvcounterH} locally renormalises the Yukawa coupling as it is expected from the standard approach:
\beq
y_q^0 \, \mu^\ep = y_q \, \mu_\uv^\ep \left( 1 - \aas C_F \frac{3}{\ep} + {\cal O}(\as^2) \right)~. 
\eeq

The vacuum amplitude correctly accounts for the derivative of the quark and antiquark selfenergies through the residues of the corresponding squared Feynman propagators. However, it also generates a term proportional to the square of the mass. This term would not be considered in the state-of-the-art approach because the renormalised selfenergy is fixed to vanish on shell. The term $\Delta Z_m^{(\uv)}$ in \Eq{eq:uvcounterZH} restores the mass renormalisation in the on-shell scheme when the renormalisation scale is identified with the quark mass itself. The requirement that $\mu_\uv=m$ in $\Delta Z_m^{(\uv)}$ also suggests an interesting physical interpretation. 

In the case of an off-shell photon, the local UV counterterm reads
\bea
&& \aduv{3,\gamma^*}(456)  =  \frac{\go{1}}{x_{45}} \bigg[\Delta Z_m^{(\uv)} (- 8 m^2) \left(1 - \frac{\beta^2}{d-2} \right)  + {\bf \Delta}_{\gamma^*}^{(\uv)} \bigg]~, 
\label{eq:auvgamma}
\eea
where $\Delta Z_m^{(\uv)}$ is the same as in \Eq{eq:uvcounterZH}, and the function that integrates to zero is 
\bea
&& {\bf \Delta}_{\gamma^*}^{(\uv)} = 
\frac{c_{\gamma}^{(\uv)}}{s \lambda_{\uv}^3}  
\left( - \frac{3(\lb_1\cdot \lb_2)^2}{\lambda_{\uv}^2} + \lb_2^2 + 2 \lb_1\cdot \lb_2 \right) 
\MM{\gamma^*\to q\bar q}{0} \nn \\ &&
\quad + \frac{3(2\lb_1\cdot \lb_2)^2}{\lambda_{\uv}^2}  + \left( d-4 + \frac{3\mu_\uv^2}{\lambda_\uv^2} -\beta^2 \right) s~, 
\eea
such that 
\bea
&&\int d\Phi_{\lb_1} \aduv{3,\gamma^*}(456) \ps{45\bar 6} = \frac{\pi}{s \beta} \, \go{1} \, \Se 
\frac{\mu^{2\epsilon}}{m^{2\epsilon}} \left( \frac{3}{\epsilon} + 4 \right) (-8m^2) \left(1 - \frac{\beta^2}{d-2} \right)~.
\label{eq:integratedUVgamma}
\eea
This result is consistent with the fact that conserved or partially conserved currents, such as the vector and axial currents, do not get renormalised. In other words, the UV counterterm in~\Eq{eq:auvgamma} renormalises the quark mass in the on-shell scheme, exactly as~\Eq{eq:uvcounterH} does, but does not renormalise the wave function because the term proportional to the squared LO is absent. Note that the vertex and selfenergy diagrams contribute with opposite signs in the UV, and therefore by considering a unitary approach there is a manifest cancellation between them without the need to subtract their UV behaviour separately. Note also that for massless quarks the local UV counterterms in \Eq{eq:uvcounterH}
and \Eq{eq:auvgamma} are well defined, in particular, the term containing $\Delta Z_m^{(\uv)}$ is proportional to the quark mass and thus has an smooth massless limit. The integrated UV counterterms in \Eq{eq:integratedUVH} and \Eq{eq:integratedUVgamma} behave as $m^2 \ln{m^2}$, while the customary wave function and mass renormalisation constants~\cite{Sborlini:2016hat} behave as $\ln{m^2}$ and are therefore ill-defined in the massless limit. This is another advantage of the LTD causal unitary approach over other implementations.

Summarising, the locally renormalised phase-space residue in~\Eq{eq:decayrateNLO} is then given by the four dimensional limit of the difference
\beq
\ad{3,f,\r}(456) = \left. \left( \ad{3,f}(456)-\aduv{3,f}(456) \right) \right|_{d=4}.
\eeq
Beyond the one loop, the construction of the local UV counterterm, although more subtle because it requires subtracting the UV behaviour of the phase-space residues in different combinations of UV configurations of the loop momenta, is algorithmically well defined in terms of the expansions defined by~\Eq{eq:uvexpansion}, as shown in detail in Ref.~\cite{Rios-Sanchez:2024xtv}.

\section{A toy decay rate at NNLO}
\label{sec:toy}

\begin{figure}[t]
\begin{center}
\includegraphics[scale=0.65]{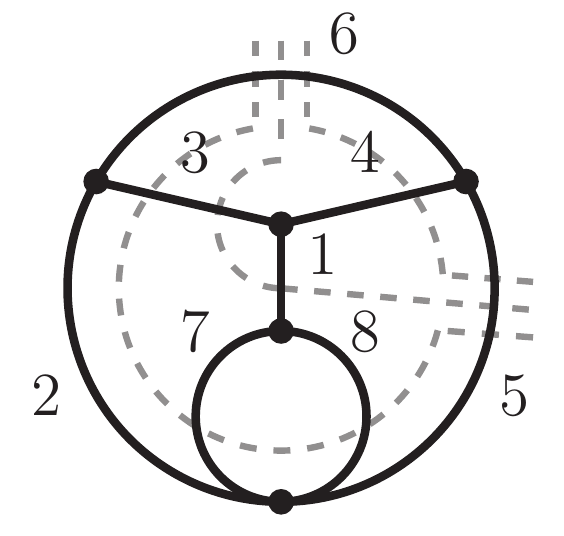}
\caption{Four-loop vacuum diagram contributing to the decay rate of a massive scalar particle at NNLO. The gray dashed lines represent phase-space residues.
\label{fig:nnlo}}
\end{center}
\end{figure}

We now consider the decay of a very heavy scalar into lighter or massless scalars at NNLO. Actually, we will not consider this decay in a realistic scalar theory, but rather in the simplified scenario generated by a single four-loop vacuum topology, which is shown in Fig.~\ref{fig:nnlo}. This simplification contains all the necessary elements to illustrate LTD causal unitary at NNLO, while allowing us to explicitly describe the two-loop, one-loop and tree-level phase-space residues with very compact expressions. Only the selfenergy-like insertion needs to be renormalised. With respect to the previous examples, which are defined by the set of momenta in~\Eq{eq:qs}, we now have two extra propagators with four-momenta
\beq
q_7=\ell_4~, \qquad q_8= \ell_4+\ell_{12}~,
\eeq
and two extra independent integration variables, the modulus~$|\lb_4|$ and the polar angle between the three-momenta~$\lb_4$ and~$\lb_{12}$. The NNLO contribution to the decay rate is
\bea
&& d\Gamma^{(2)}_{\Phi\to \phi\phi} = \frac{d\Phi_{\lb_1\lb_2\lb_4}}{2\sqrt{s}} \, \bigg[
\Big( \ad{4,\Phi,\r}(456) \, \ps{45\bar 6}  \nn \\ 
&& \quad + \ad{4,\Phi,\r}(1356) \, \ps{135\bar 6} 
+ \ad{4,\Phi}(35678) \, \ps{35\bar 6 78} \Big) + (5\leftrightarrow 2, 4\leftrightarrow 3) \bigg]~.
\label{eq:decayrateNNLO}
\eea

The phase-space residue involving two-loop amplitudes is  
\bea
&& \ad{4,\Phi}(456) = \frac{g_\Phi^{(2)}s^3}{x_{1234578}}
\Bigg[ \frac{1}{\lambda_{13\bar 4}}\left(
L^{23\bar 4\bar 5}_{3\bar 4 78, 2\bar 5 78} + L^{2578}_{125, 3\bar 4 78}
+ L^{178}_{125, 2\bar 5 78} \right) \nn \\ &&
\quad + \frac{1}{\lambda_{134}}\left( 
L^{2578}_{2345, 178} + L^{3478}_{2345, 2\bar 5 78} \right) 
+ \frac{1}{\lambda_{12\bar 5}} \left( 
L^{134}_{2\bar 5 78, 178}+L^{23\bar 4\bar 5}_{2\bar 5 78, 178}
\right)
+ \frac{1}{\lambda_{125}} L^{2345}_{2578, 178} \nn \\ &&
\quad + \frac{1}{\lambda_{178}}\left( 
L^{3478}_{2345, 2\bar 5 78} + L^{3\bar 478}_{23\bar 4\bar 5, 2578} \right) \Bigg]~,
\eea
with
\beq
L^i_{j,k} = \frac{1}{\lambda_i} \left(
\frac{1}{\lambda_j} + \frac{1}{\lambda_k} \right)~.
\eeq
There is another phase-space residue at one-loop
\bea
&& \ad{4,\Phi}(1356)  =  \frac{g_\Phi^{(2)}s^3}{x_{13578}}
\left(\frac{1}{\lambda_{13 \bar 4} \lambda_{134} \lambda_{1\bar 2 5} \lambda_{125}} \right) 
\left(\frac{1}{\lambda_{\bar 1 78}} + \frac{1}{\lambda_{178}}
\right)~,
\eea
and the tree-level contribution 
\bea
&&\ad{4,\Phi}(35678) = - \frac{g_\Phi^{(2)}s^3}{x_{3578}}   \left(\frac{1}{
\lambda_{3 \bar 4 78} \lambda_{3478}
\lambda_{\bar 2 578} \lambda_{2578}
\lambda_{\bar 1 78} \lambda_{178}}
\right)~. 
\eea
In addition, we should consider those phase-space residues that are obtained by exchanging the indices $(2\leftrightarrow 5, 3\leftrightarrow 4)$, i.e. $\ad{4,\Phi}(236)$, $\ad{4,\Phi}(1246)$ and~$\ad{4,\Phi}(24678)$. Notice that these expressions are the same regardless of whether the final-state scalars are massive or massless since the scalar masses are implicit in the on-shell energies. 

It is then easy to realise how collinear and quasicollinear singularities cancel locally among different phase-space residues. For example, there are two double-collinear singularities at  $\lambda_{13\bar 4} \to~0$ and $\lambda_{\bar{1}78}\to 0$, respectively, which cancel out as
\bea
&& \lim_{\lambda_{13\bar 4} \to 0} 
\left( \ad{4,\Phi}(456) \ps{45\bar 6} + 
\ad{4,\Phi}(1356) \ps{135\bar 6} 
\right)  = {\cal O}(\lambda_{13\bar 4}^0)~, \nn \\
&& \lim_{\lambda_{\bar 1 78} \to 0} 
\left( \ad{4,\Phi}(1356) \ps{135\bar 6} + 
\ad{4,\Phi}(35678) \ps{35\bar 6 78} 
\right)  = {\cal O}(\lambda_{\bar 1 78}^0)~.
\label{eq:doublennlo}
\eea
When both double-collinear singularities occur simultaneously, which is equivalent to the limit $\lambda_{3\bar{4}78}\to 0$, a triple-collinear singularity emerges, and the three phase-space residues are needed to achieve a local cancellation
\bea
&& \lim_{\lambda_{3\bar 4 78} \to 0} 
\left( \ad{4,\Phi}(456) \ps{45\bar 6} + \ad{4,\Phi}(1356) \ps{135\bar 6} 
+ \ad{4,\Phi}(35678) \ps{35\bar 6 78} \right) = {\cal O}(\lambda_{3\bar 4 78}^0)~. \nn \\
\label{eq:triplennlo}
\eea
Finally, a threshold singularity arises at $\lambda_{23\bar 4\bar 5}\to 0$ that matches locally between $\ad{4,\Phi}(456)$ and $\ad{4,\Phi}(236)$.

The two- and one-loop phase-space residues need to be renormalised in the UV region. Following the same procedure explained above, the UV counterterms of these  phase-space residues are given by 
\bea
&& \aduv{4,\Phi}(456) = \frac{g_\Phi^{(2)}s}{4\lambda_{\uv}^3} \ad{3,\Phi}(456)~, 
\qquad \aduv{4,\Phi}(1356) = \frac{g_\Phi^{(2)}s}{4\lambda_{\uv}^3} \ad{3,\Phi}(1356)~,
\eea
respectively, where the UV on-shell energy is $\lambda_\uv=\sqrt{\lb_4^2+\mu_\uv^2-\ii}$, and $\ad{3,\Phi}(456)$ and $\ad{3,\Phi}(1356)$ are the coupling-stripped phase-space residues to one less order
\bea
&& \ad{3,\Phi}(456)  =  \frac{s^2}{x_{12345}}
\left(L^{13\bar 4}_{23\bar 4 \bar 5, 125} +
L^{12\bar 5}_{23\bar 4\bar 5, 134} + 
L^{2345}_{134,125} \right)~, \nn \\ 
&& \ad{3,\Phi}(1356)  =  \frac{s^2}{x_{135}}
\left(\frac{1}{\lambda_{13 \bar 4} \lambda_{134} \lambda_{1\bar 2 5} \lambda_{125}} \right)~.
\eea
For reference, the coupling-stripped phase-space residue at~LO is dimensionless and is given by $\ad{2,\Phi}(456)= s/x_{45} = 1$, while
\bea
d\Gamma^{(1)}_{\Phi\to \phi\phi} &=& \frac{d\Phi_{\lb_1\lb_2}}{2\sqrt{s}} \, \bigg[
\Big( \ad{3,\Phi}(456) \, \ps{45\bar 6}  + \ad{3,\Phi}(1356) \, \ps{135\bar 6} \Big) 
+ (5\leftrightarrow 2, 4\leftrightarrow 3) \bigg]~.
\label{eq:decayratescalarNLO}
\eea

\section{Numerical implementation}
\label{sec:numeric}

\begin{figure}[t]
\begin{center}
\includegraphics[scale=1.1]{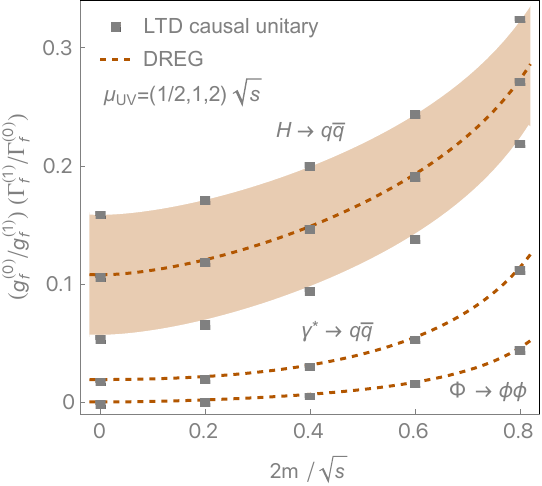}
\caption{Numerical implementation of LTD causal unitary at NLO for the three decay processes $H\to q\bar q$, $\gamma^*\to q\bar q$ and $\Phi\to \phi\phi$ as a function of the final state mass. The photon current is conserved and the scalar decay is UV finite. Consequently, they do not exhibit any dependence on the renormalisation scale $\mu_\uv$.
\label{fig:nlo}}
\end{center}
\end{figure}

With the expressions presented in the previous sections, we now introduce a numerical implementation directly in the three physical spacial dimensions, $d-1=3$, since IR singularities cancel locally among phase-space residues and the local UV counterterm properly accounts for the singular UV behaviour. Threshold singularities also match locally in the sum of the loop phase-space residues rendering the integrand well behaved across thresholds. To this end, we must consider the constraints introduced by energy conservation, which are encoded by the functions $\ps{i_1\cdots i_n\bar 6}$ in \Eq{eq:decayrateLO}, \Eq{eq:decayrateNLO} and \Eq{eq:decayrateNNLO}.

\begin{table}[t]
\begin{tabular}{|clllll|} \hline
$2m/\sqrt{s} = $ & $10^{-3}$ & $0.2$ & $0.4$ & $0.6$ & $0.8$ \\ \hline
$\Phi \to \phi \phi$ 
& $0$ & $0.001760$ & $0.006509$ & $0.016693$ & $0.045894$ \\
& $0.000029(14)$ &  $0.001785(10)$ & $0.006517(7)$ & $0.016702(8)$ & $0.045929(10)$ \\ \hline
$\gamma^* \to q\bar q(g)$ & $0.018998$ & $0.021524$ & $0.031273$ & $0.054715$ &  $0.114001$ \\
& $0.018963(57)$ & $0.021556(24)$ & $0.031280(22)$ & $0.054785(36)$ & $0.114013(19)$ \\ \hline
$H\to q\bar q (g)$ & $0.054981$ &  $0.067702$ &  $0.095937$ & $0.140174$ & $0.220367$ \\
$\mu_\uv=2 m_H$ 
& $0.055017(33)$ & $0.067713(17)$ & $0.095948(16)$ & $0.140200(16)$ & $0.220364(16)$ \\ \hline
$H\to q\bar q (g)$ & $0.107654$ & $0.120374$ & $0.148610$ & $0.192847$ & $0.273040$ \\
$\mu_\uv=m_H$ & $0.107642(45)$ & $0.120361(24)$ & $0.148629(23)$ & $0.192858(29)$ & $0.273013(24)$ \\ \hline
$H\to q\bar q (g)$ & $0.160327$ & $0.173047$ & $0.201283$ & $0.245520$ & $0.325713$ \\
$\mu_\uv=m_H/2$ & $0.160305(56)$ & $0.173070(50)$ & $0.201319(45)$ & $0.245522(44)$ & $0.325708(41)$ \\ \hline
\end{tabular}
\caption{Numerical implementation of LTD causal unitary at NLO for the three decay processes $H\to q\bar q$, $\gamma^*\to q\bar q$ and $\Phi\to \phi\phi$ as a function of the final state mass. The first row in each item is the analytical result in DREG from the expressions presented in Appendix~\ref{app:dreg}. The photon current is conserved and the scalar decay is UV finite. Consequently, they do not exhibit any dependence on the renormalisation scale $\mu_\uv$. \label{table:nlo}}
\end{table}
The phase-space energy function $\ps{45\bar 6}$ only involves the $\lb_2$ and $\lb_3$ loop momenta  and leaves $\lb_1$ unconstrained. Since at the centre-of-mass frame we set $\lb_3=\boldsymbol{0}$, then $\ps{45\bar 6}$ fixes the modulus of $\lb_2$ and leaves the angular dependence unconstrained. Conversely, $\ps{23\bar 6}$ fixes the modulus of $\lb_1$. This is the expected constraint of a two-body phase space. The phase-space energy functions $\ps{135\bar 6}$ and $\ps{124\bar 6}$ involve the three loop momenta. Since all the loop momenta are now bounded by $\qon{6} = \sqrt{s-\ii}$, then the integration domain is restricted to a compact region of the phase space, corresponding to a three-body phase space. 

If $p_1$ and $p_2$ are the momenta of the quark and the antiquark, respectively, in the final-state with two particles, and $p_1'$, $p_2'$ and $p_3'$ are the momenta of the quark, antiquark and gluon in the final state with three particles, the following linear mappings apply 
\bea
&& p_a(6) \to p_1(4) + p_2(\bar 5)~, \nn \\
&& p_a(6) \to p_1(\bar 3) + p_2(2)~, \nn \\
&& p_a(6) \to p_1'(4) + p_2'(2) + p_3'(\bar 1)~, \nn \\
&& p_a(6) \to p_1'(\bar 3) + p_2'(\bar 5) + p_3'(1)~,
\eea
where the mapping is interpreted in the following sense
\beq
p_s  (i): p_s^\mu = (\sqrt{\qb_{i}^2+m_{i}^2-\ii}, \qb_{i})~,
\eeq 
and the on-shell energies are 
\bea
&& \qon{1} =  \sqrt{\lb_{12}^2-\ii}~, \nn \\
&& \qon{2} =  \qon{3} = \sqrt{\lb_1^2+m^2-\ii}~, \nn \\
&& \qon{4} =  \qon{5} = \sqrt{\lb_2^2+m^2-\ii}~.
\eea
As expected, there are five independent integration variables considering two modulus, four angles and one energy conservation constraint. Nevertheless, assuming the decaying particle to be at rest the number of effective independent variables is two.

\begin{figure}[t]
\begin{center}
\includegraphics[scale=1.1]{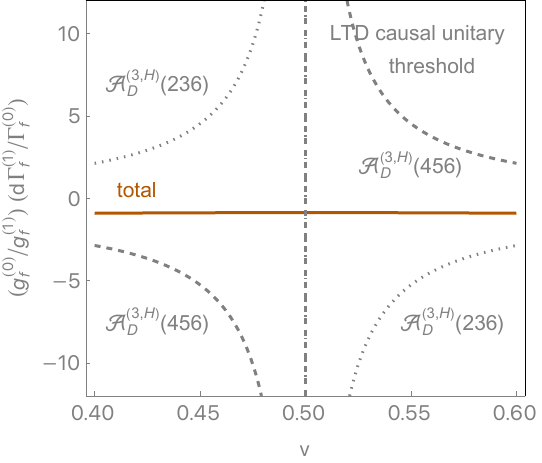}
\caption{Unintegrated decay rate across a threshold singularity as a function of the angular variable. The sum over the one-loop phase-space residues is flat. 
\label{fig:threshold}}
\end{center}
\end{figure}

\begin{figure}[t]
\begin{center}
\includegraphics[scale=1.1]{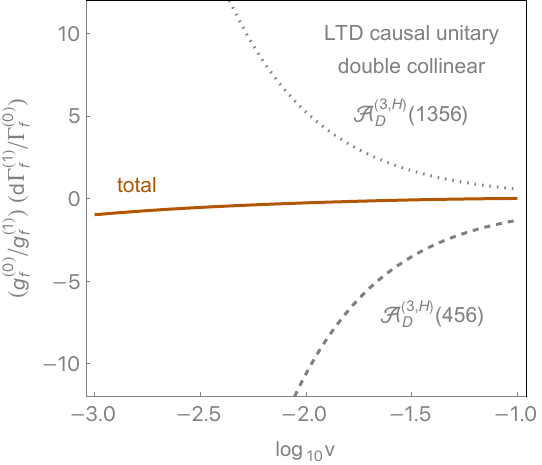}
\caption{Local cancellation of collinear singularities at NLO between phase-space residues with different numbers of final-state particles. 
\label{fig:collinear}}
\end{center}
\end{figure}

The phase-space energy function $\ps{35\bar 6 78}$ in \Eq{eq:decayrateNNLO} and its symmetric counterpart $\ps{24\bar 6 78}$ introduce a dependence in the $\lb_4$ loop momentum, and therefore two additional independent integration variables. We should also consider final-states with four external particles and the linear mappings
\bea
&& p_a(6) \to p_1''(4) + p_2''(2) + p_4''(\bar 7) + p_5''(\bar 8)~, \nn \\
&& p_a(6) \to p_1''(\bar 3) + p_2''(\bar 5) + p_4''(7) + p_5''(8)~,
\eea
with
\bea
&& \qon{7} =  \sqrt{\lb_{4}^2+m^2-\ii}~, 
\qquad \qon{8} = \sqrt{(\lb_4+\lb_{12})^2+m^2-\ii}~.
\eea

The numerical implementation is quite stable, in particular due to the absence of threshold singularities. The numerical results obtained for the NLO contribution to the decay rates, as a function of the final-state mass and normalised to the LO total decay rate, are shown in Fig.~\ref{fig:nlo}, where they are compared with the state-of-the-art DREG analytical expressions (see Appendix~\ref{app:dreg}). The corresponding numerical values are presented in Table~\ref{table:nlo}. We have used a private version of VEGAS~\cite{Lepage:1977sw} which achieves a precision of $10^{-3}-10^{-4}$ with $10$ iterations of $10^7$ calls. This is a few seconds per point on a personal laptop. Only $H\to q\bar q$ exhibits a dependence on the renormalisation scale $\muUV$ because the vector current of the photon decay is UV protected and the scalar decay is UV finite at NLO. Similar results have been obtained by using Quantum Fourier Iterative Amplitude Estimation~(QFIAE)~\cite{deLejarza:2023qxk,deLejarza:2024pgk}. This is a quantum algorithm for numerical integration of multidimensional functions that trains a quantum neural network to decompose the target function into its Fourier components, and then takes advantage of the fact that trigonometric functions are efficiently integrable in a quantum approach. We refer the interested reader to Ref.~\cite{deLejarza:2024scm}. 

In Fig.~\ref{fig:threshold} and Fig.~\ref{fig:collinear}, we illustrate the integrand behaviour of the different phase-space residues and their sum across a threshold singularity and in collinear configurations, respectively. We can clearly observe how the different phase-space residues are singular, but their sum provides a flat integrand. The phase-space residues shown in Fig.~\ref{fig:threshold} and Fig.~\ref{fig:collinear} correspond to $H\to q\bar q$. Similar results are obtained for the other two processes. 

\begin{figure}[t]
\begin{center}
\includegraphics[scale=1.1]{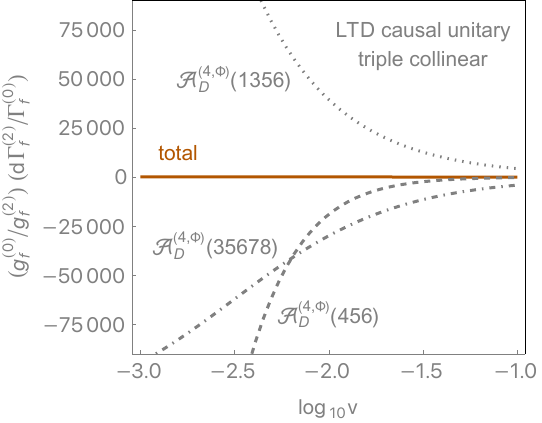}
\caption{Local cancellation of collinear singularities at NNLO between phase-space residues with different numbers of final-state particles. 
\label{fig:triplecollinear}}
\end{center}
\end{figure}

The local cancellation of collinear singularities at NNLO is illustrated in Fig.~\ref{fig:triplecollinear} for the toy model described in Section~\ref{sec:toy}. We have fixed the angular variable that defines the collinearity of particles $7$ and $8$ to $v_4=10^{-2}$. We can observe that for large values of the other angular variable, $v$, the  cancellation occurs mostly between $\ad{4,\Phi}(1356)$ and $\ad{4,\Phi}(35678)$, which corresponds to the local cancellation of the double-collinear singularity at $\lambda_{\bar 1 78} \to 0$ according to \Eq{eq:doublennlo}. At very small angles, $v < v_4$, instead $\ad{4,\Phi}(456)$ becomes much more singular than $\ad{4,\Phi}(35678)$, reflecting the dominance of the emergent triple-collinear configuration, and the participation of the three phase-space residues is necessary for the local cancellation of the overlapping collinear singularities, according to \Eq{eq:triplennlo}. 

Comparing Fig.~\ref{fig:triplecollinear} and Fig.\ref{fig:collinear}, we also observe that the local cancellation of collinear singularities is equally effective at NLO and NNLO, although the scale of the individual singularities is, as expected, much larger at NNLO. 

\section{Conclusions}
\label{sec:conclusions}

We have presented the first proof-of-concept implementation of LTD causal unitary~\cite{Ramirez-Uribe:2024rjg} to decay processes at higher perturbative orders. The processes chosen are such that all the benefits of the method are clearly visible in a lightweight manner. Specifically, the central and novel ingredient of LTD causal unitary is the use of a multiloop vacuum amplitude in the LTD representation as a kernel of all the states contributing to the decay process. The LTD representation of the vacuum amplitude, which is manifestly causal, coherently encodes the final states as residues on causal propagators that depend on linear combinations of the on-shell energies of the internal propagators, once certain on-shell energies identified with the incoming particles are analytically continued to negative values. 

LTD causal unitary ensures the local cancellation of soft and collinear or quasicollinear singularities to all orders in perturbation theory through the unitary sum of the phase-space residues, while the renormalisation of UV singularities is achieved through suitable UV local counterterms. The use of vacuum amplitudes as a kernel incorporates the wave-function renormalisation of external particles in a well-defined approach that is free from mathematical ambiguities. The mass renormalisation is also physically interpreted as a function of a proper renormalisation scale. 

Concerning threshold singularities, which typically represent an overwhelming obstacle in numerical implementations, LTD causal unitary provides an efficient solution leading to flat integrands across thresholds, as the potential singularities match between phase-space residues with the same number of final-state particles. This property, which is a natural consequence of using a vacuum amplitude as a kernel, represents a clear advantage.

The local cancellation of singularities at NLO and NNLO has been illustrated with selected decay processes. The total decay rates at NLO have been obtained with LTD causal unitary and have been compared with state-of-the-art DREG analytical expressions, showing a perfect agreement. The transition from a massive to a massless implementation, as first observed in \cite{Sborlini:2016hat}, is smooth. The numerical implementation in LTD causal unitary is quite stable and leads to accurate results with minimal CPU resources.

The results presented in this paper constitute a solid confirmation of the unique capabilities and advantages of LTD causal unitary at higher perturbative orders, in addition to successfully addressing the original motivation and path initiated with the seminal works on the loop-tree duality~\cite{Catani:2008xa,Bierenbaum:2010cy,Bierenbaum:2012th}. A proof-of-concept implementation for scattering processes that require a local subtraction of initial-state collinear singularities, as well as more complex and well-motivated phenomenological analysis with LTD causal unitary for scattering and decay processes at high-energy colliders, will be presented in forthcoming publications.

\section*{Acknowledgements}

This work is dedicated to Stefano Catani, who always inspired us in the development of this research line.  Part of this work was completed during the Workshop {\it Theory Challenges in the Precision Era of the Large Hadron Collider} held at the Galileo Galilei Institute (Firenze) in September 2023. This work is supported by the Spanish Government (Agencia Estatal de Investigaci\'on MCIN/ AEI/10.13039/501100011033) Grants No.~PID2020-114473GB-I00, No. PID2022-141910NB-I00, No. PID2023-146220NB-I00 and No. CEX2023-001292-S, and Generalitat Valenciana Grant No. PROMETEO/2021/071. AERO is supported by the Spanish Government (PRE2018-085925). DFRE and JML are supported by Generalitat Valenciana (CIGRIS/2022/145 and ACIF/2021/219). PKD is supported by European Commission MSCA Action COLLINEAR-FRACTURE, Grant Agreement No. 101108573. SRU and RJHP are funded by CONAHCyT through Project No.~320856 (Paradigmas y Controversias de la Ciencia 2022), Ciencia de Frontera 2021-2042 and Sistema Nacional de Investigadores. LC is supported by Generalitat Valenciana GenT Excellence Programme (CIDEGENT/2020/011) and ILINK22045. GS is partially supported by EU Horizon 2020 research and innovation programme STRONG-2020 project under Grant Agreement No. 824093 and H2020-MSCA-COFUND USAL4EXCELLENCE-PROOPI-391 project under Grant Agreement No 101034371. WJT is supported by the Leverhulme Trust, LIP-2021-01.

\appendix
\section{The phase space}
\label{app:phasespace}

Our representation of the phase space for $m$ particles in the final state is derived from the standard expression
\beq
d\Phi_m =  \mu^{d-4} (2\pi)^d \, \delta^{(d)}\left( \sum_{i=1}^m p_i - p_{ab} \right) \prod_{i=1}^m \frac{\mu^{4-d} \, d^{d-1}p_i}{(2\pi)^{d-1} 2E_i}~,
\eeq
where $p_{ab}=  p_a+p_b$ stands for the sum of initial-state momenta. Inspired by the LTD representation of multiloop amplitudes, we first define
\beq
d\Phi_{\pb_i} = \mu^{4-d} \frac{d^{d-1} p_i}{(2\pi)^{d-1}}~, 
\eeq
which is typically parametrised as
\bea
&& d\Phi_{\pb_i} = \frac{\mu^{2\epsilon}}{(2\pi)^{d-1}} (\pb_i^2)^{1-\epsilon} d|\pb_i| \, d\Omega_i^{(d-2)}~,  \nn \\
&& d\Omega_i^{(d-2)} = \frac{(4\pi)^{1-\epsilon}}{\Gamma(1-\epsilon)} (v_i (1-v_i))^{-\epsilon} dv_i~.
\eea
Then, we identify $E_i \to p_{i,0}^{(+)} = \sqrt{{\bf p}_i^2+m_i^2- \ii}$ as the on-shell energies, to rewrite the phase space as
\bea
d\Phi_m &=& \mu^{d-4} \, (2\pi)^{d-1}\, \delta^{(d-1)}\left( \sum_{i=1}^m \pb_i - \pb_{ab} \right) \,   \frac{\ps{1\cdots m \bar a\bar b}}{x_{1 \cdots m}} \, d\Phi_{{\bf p}_1, \ldots, {\bf p}_n} ~,
\eea
where $x_{1\cdots m} = \prod_{i=1}^m 2 p_{i,0}^{(+)}$ is the product of all the on-shell energies of the final-state particles. The energy-conservation Dirac delta function is specifically given by 
\beq
\ps{1\cdots m \bar a\bar b} =  2\pi \,  \delta(\lambda_{1\cdots m \bar a\bar b})~, 
\eeq
where
\beq
\lambda_{1\cdots m \bar a \bar b} = \sum_{i=1}^m p_{i,0}^{(+)} - p_{a,0}^{(+)} - p_{b,0}^{(+)}~.
\eeq

\section{DREG analytical expressions for NLO decay rates}
\label{app:dreg}

The following expressions for the NLO predictions of the decay rates from DREG are adapted from Ref.~\cite{Sborlini:2016hat}.
First, we define the function
\beq
h_p = \frac{4}{\beta} \left[ \ln{x_S} 
\left( \ln{1-x_S} + \ln{1+x_S^2} \right) + 2\li{x_S} + \li{x_S^2} 
\right]~,
\eeq
where $x_S=(1-\beta)/(1+\beta)$, with $\beta = \sqrt{1-4m^2/s}$. We also define the functions
\bea
{\cal F}_H &=& (1+\beta^2) h_p - 2 \ln{\frac{m^2}{s}} - 4 \left( \frac{1-2\beta}{\beta} \ln{x_S} + 4 \ln{1-x_S} -5\right)~, \nn \\ 
{\cal F}_{\gamma^*} &=& (1+\beta^2) h_p - 2 \ln{\frac{m^2}{s}} \nn \\ &-& 4 \left( \frac{2-7\beta}{2\beta} \ln{x_S} + 4 \ln{1-x_S} + 3 \ln{1+x_S}- 3\right)~,
\eea
and
\bea
&& {\cal G}_H = -3 \left(1+\beta^2\right) + \frac{13\beta^4-18\beta^2-3}{2\beta} \ln{x_S}~, \nn \\
&& {\cal G}_{\gamma^*} = 3 \left(\beta^2-7\right) + \frac{7\beta^4-30\beta^2-9}{2\beta} \ln{x_S}~.
\eea
The coupling striped theoretical prediction for the decay rates at NLO, normalized to the LO, is 
\beq
{\bf \Gamma}_f^{(1)} = 
\frac{g_f^{(0)}}{g_f^{(1)}} 
\frac{\Gamma_f^{(1)}}{\Gamma_f^{(0)}}~,
\eeq
with 
\bea
&& {\bf \Gamma}_\Phi^{(1)} = \frac{h_p}{16\pi^2}~, \nn \\
&& {\bf \Gamma}_H^{(1)} = \frac{1}{16\pi^2}
\left[ 6 \ln{\frac{s}{\mu_\uv^2}} + {\cal F}_H + \frac{{\cal G}_H}{2\beta^2} 
\right]~, \nn \\
&& {\bf \Gamma}_{\gamma^*}^{(1)} = \frac{1}{16\pi^2}
\left[ {\cal F}_{\gamma^*} + \frac{{\cal G}_{\gamma^*}}{3-\beta^2} 
\right]~.
\eea
Only the theoretical prediction for the Higgs boson decay has a dependence on the renormalisation scale $\mu_\uv$.

\bibliographystyle{JHEP}
\bibliography{LTD_causal_unitary_timelike}

\end{document}